# Analysis of the non-Markovianity for electron transfer reactions in an oligothiophene-fullerene heterojunction

E. Mangaud[1*], C. Meier[1†] and M. Desouter-Lecomte[2,3‡]

[1] *Laboratoire Collisions, Agrégats, Réactivité, UMR 5589, IRSAMC, Université Paul Sabatier, F-31062 Toulouse, France*

[2] *Laboratoire de Chimie Physique, Bât 349, Univ Paris-Sud, UMR 8000, Université Paris Saclay F-91405 Orsay, France*

[3] *Département de Chimie, Université de Liège, Sart Tilman, B6, B-4000 Liège, Belgium*

**ABSTRACT**

The non-Markovianity of the electron transfer in an oligothiophene-fullerene heterojunction described by a spin-boson model is analyzed using the time dependent decoherence canonical rates and the volume of accessible states in the Bloch sphere. The dynamical map of the reduced electronic system is computed by the hierarchical equations of motion methodology (HEOM) providing an exact dynamics. Transitory witness of non-Markovianity is linked to the bath dynamics analyzed from the HEOM auxiliary matrices. The signature of the collective bath mode detected from HEOM in each electronic state is compared with predictions of the effective mode extracted from the spectral density. We show that including this main reaction coordinate in a one-dimensional vibronic system coupled to a residual bath satisfactorily describes the electron transfer by a simple Markovian Redfield equation. Non-Markovianity is computed for three inter fragment distances and compared with *a priori* criterion based on the system and bath characteristic timescales.

[*] *etienne.mangaud@irsamc.ups-tlse.fr*
[†] *chris@irsamc.ups-tlse.fr*
[‡] *michele.desouter-lecomte@u-psud.fr*

# I. INTRODUCTION

Electron transfer in complex molecular systems has been described for many decades by the open quantum system theory (OQS) [1, 2, 3]. The basic model is the spin-boson (SB) Hamiltonian [4, 5, 6] in which the two states of the electronic system (called spin) are coupled to a bosonic bath of vibrational oscillators. The OQS dynamics has been treated by a multitude of computational methods such as the Feynman-Vernon path integral formalism [7] leading to the hierarchical equations of motion (HEOM) [8] or time non-local master equations with perturbative approximation [9] coming from the Nakajima-Zwanzig projection technique [10] or time local methods [6, 11] from the Hashitsume expansion [12]. Alternative methods are the stochastic approach [13, 14], density matrix renormalization group [15] or the simulation of the bath by a finite number of oscillators treated by multi configuration time dependent Hartree (MCTDH) [16, 17] with the multi-layer implementation (ML-MCTDH) simulations [18] or the time-dependent variational matrix product states [19].

The physical realization and the control [20] of OQS play an essential role for future quantum technology. A still open question is the role of the electronic coherence protection during the transfer and in particular, the role of the information back flow from the dissipative bath to the system. This question is directly linked to the memory effects or in other words to the non-Markovianity of the dissipative dynamics. The partition of the two-state system from its surrounding leads to a reduced equation for the system density matrix containing a memory integral in which the bath interplays through an autocorrelation function of a collective bath mode. A correlation function that decays much more rapidly than the characteristic timescale of the system, here the Rabi period of the electronic motion, leads to memoryless Markovian dynamics. However, in many applications of electron or excitation transfer in organic macromolecules, photovoltaic materials, cryptochromes and DNA photolyases or nanostructured protein complexes of photosynthetic organisms, the bath correlation time is not negligible and the Markov approximation breaks down. The relaxation and decoherence of the electronic system can be seen as a transfer of information between the electronic and vibrational degrees of freedom. This information flow is unidirectional in a Markovian approach but non-Markovianity allows for some flow back to the electronic system. Non-Markovianity could be of great interest in many situations for instance for precision estimation under noise in quantum metrology [21, 22] in protocols for teleportation [23], steady-state entanglement maintenance [24, 25] or for transport enhancing in biopolymers [26].

Many factors can influence the non-Markovianity, such as the temperature, the initial system-bath entanglement and the structure of the environment. The latter is mainly described by the spectral density that gives the system-bath coupling as a function of the frequency. Different questions arise: What is the importance of being on or off-resonance with some peaks in the spectral density? What is the role of the coupling intensity? Can we identify the main collective mode capturing the system-bath correlation, a mode which could be included in the system providing an effective vibronic model weakly coupled to a residual Markovian bath? The crucial role of a vibrational mode enhancing the electronic coherence has been addressed recently [27, 28, 29] and examined from the non-Markovianity view point for the excitation energy transfer in pigment-proteine complexes [30]. In order to understand the relation between the structure of the environment and the non-Markovianity and to quantity these effects, several measures have been proposed recently. A review can be found in references [31], [32], [33] and [34]. The measure used by Breuer [35] is based on the quantum trace distance measuring the distinguishability between two quantum states. A temporarily increase of this trace distance is taken as a signature of backflow of information during the dissipative evolution. Another measure estimates the volume of accessible states in the Bloch sphere [36, 37]. Non-Markovianity is then detected by a temporary increase of this volume while a monotonous decrease is expected in the Markovian case. Another interesting approach analyzes the canonical relaxation rate constants [31] that are constant in a Markovian process and calibrate a canonical Lindblad Master equation [38, 39]. The rates become time dependent and some can be temporarily negative in a non-Markovian evolution. All these measures allow one to detect the crossover between Markovian and non Markovian regimes.

Electron transfer is studied here in the context of organic photovoltaic (OPV) device for which there has been many theoretical works to interpret the ultrafast charge separation [40, 41, 42, 43, 44]. The main nowadays speculation concerns the ultimate conception of new materials inspired by the efficient coherent transport in photosynthetic systems [45, 46, 47] The example is here the charge exchange between the photoinduced excitonic donor state $OT_4^*$-$C_{60}$ denoted XT and a charge separated state $OT_4^+ - C_{60}^-$ denoted CT that has already been studied by different approaches. Electronic dynamics has first been treated by the MCTDH (Multi Configuration Time Dependent Hartree) method with an approximate model Hamiltonian [48, 49, 50, 51, 55, 56] by sampling the oscillator bath formed by all the normal modes of both fragments. Then dissipative dynamics has been carried out by HEOM with a hierarchical decomposition of the spectral density [52] or by the auxiliary matrix method in a perturbative treatment of the vibronic model obtained by including one effective bath mode in the system Hamiltonien [53]. The vibronic hot nature of the charge transfer has also been considered in a multi-state model [54]. The present system is particularly appealing since the inter fragment distance can be taken as a variable parameter scanning very different situations of intersystem electronic coupling and gap and therefore of Rabi frequency. The inter fragment mode may be considered in a first approach as a spectator mode weakly coupled to the relaxation and decoherence dynamics. The role of the inter fragment coordinate in the electron transfer is analyzed in reference [56]. The bath spectral density is then structured by all the other vibrational modes and is characterized by a single correlation time for all the inter fragment distances. By varying the Rabi frequency, one modifies the detuning of the system frequency with respect to the main peaks in the spectral density and the ratio between the bath correlation time and the typical timescale of the electronic system.

In this paper, the dissipative dynamics is treated by the HEOM method that requires an efficient decomposition of the correlation function into a set of complex exponential functions associated with dissipation modes. This corresponds to a particular parametrization of the spectral density. Different possibilities have been considered in the literature and in particular the Tannor-Meier [9] parametrization in terms of two-pole Lorentzian functions leading to an Ohmic behavior for small frequencies. Here, we want to model a heterojunction in a solid phase in which a super Ohmic behavior is more expected. Therefore, we extend the formalism to use four-pole Lorentzian functions to fit the spectral density (see Appendix for details).

The paper is organized as follows. The spin-boson (SB) model and its parametrization are presented in section II. Section III summarizes the HEOM method and section IV presents the non-Markovianity measures. The electron transfer dynamics and its non-Markovianity is analyzed in Section V. The possible information backflow revealed by the measure is correlated with some signature of the bath dynamics obtained from the HEOM auxiliary matrices. This suggests an interesting comparison with the effective one-dimensional vibronic model which is presented in section VI. Finally, section VII concludes.

## II SB MODEL OF THE HETEROJUNCTION

### A. Molecular system

Exciton dissociation is a basic mechanism in the energy conversion in organic photovoltaic cells. The typical timescale of this process is a few hundreds of femtoseconds. The exciton dissociation dynamics of an oligothiophene ($OT_4$)/fullerene ($C_{60}$) donor-acceptor complex has been recently investigated and calibrated by *ab initio* calculations [55, 56]. The fragments are represented in Figure 1. It is a model for bulk heterojunctions of conjugated polymers and fullerene (poly-3-hexylthiophene (P3HT) and phenyl-$C_{61}$ butyric acid methyl ester (PCBM)).[57, 58]

We base our analysis on the *ab initio* data of the two XT and CT electronic diabatic states computed by long-range corrected TDDFT (see Ref. [59] for details on the diabatization procedure). The inter-fragment ($OT_4$-fullerene) distance ($R$) is not taken as a dynamical coordinate for the electronic transfer but is a parameter modulating the energy gap between the diabatic states and the electronic coupling [56]. These *ab initio* data and the corresponding Rabi period are given for three inter fragment distances in Table 1.

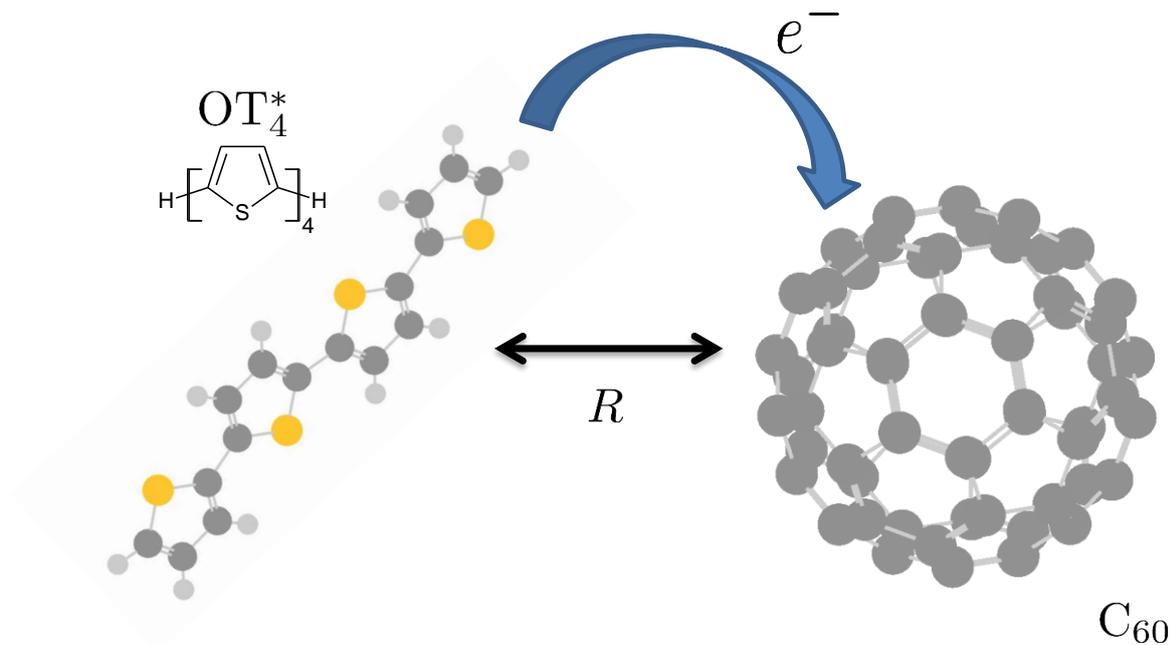

FIG.1. Model of an oligothiophene ($OT_4$) –fullerene ($C_{60}$) heterojunction. *R* corresponds to different intermolecular distances of the two fragments.

TABLE 1. Energy gap Δ between the diabatic XT and CT states and electronic coupling $V$ for different inter fragment distances $R$.

| $R$ (Å) | $\Delta_{XT\text{-}CT}$ (eV) | $V_{XT\text{-}CT}$ (eV) | $\tau_{Rabi}$ (fs) | $\Omega_{Rabi}$ (cm$^{-1}$) |
|---|---|---|---|---|
| 2.50 | 0.517 | 0.200 | 6.3 | 4998 |
| 3.00 | 0.210 | 0.130 | 12.3 | 2421 |
| 3.50 | 0.007 | 0.070 | 29.4 | 856 |

H. Tamura *et al.* [55] also carried out a normal mode analysis for the different fragments (OT$_4^+$-C$_{60}^-$) and determined the displacements $d_i$ between the minima of the two states XT and CT for the $M = 264$ modes that lead to the vibronic coupling constants in the SB model described below. This last step is performed only once assuming that the deformations will be similar whatever the inter fragment distance $R$.

## B. Spin-Boson model

The system Hamiltonian is reduced to the diabatic coupled electronic states at a given inter fragment distance $R$. The environment is modeled by a bath of harmonic oscillators of frequency $\omega_i$ for the normal coordinate $\bar{q}_i$. Assuming that the geometrical deformations implied by the charge transfer are small, one can describe them using the same oscillators in both states but with different displacements $d_i$ on the charge transfer state leading to the following Hamiltonian:

$$\hat{H} = \begin{pmatrix} 0 & V_{XT-CT} \\ V_{XT-CT} & \Delta_{XT-CT} \end{pmatrix} + \begin{pmatrix} \frac{1}{2}\sum_{i=1}^{M} \bar{p}_i^2 + \omega_i^2 \bar{q}_i^2 & 0 \\ 0 & \frac{1}{2}\sum_{i=1}^{M} \bar{p}_i^2 + \omega_i^2 (\bar{q}_i - d_i)^2 \end{pmatrix}. \quad (1)$$

After the change of coordinates: $q_i = \bar{q}_i + \frac{d_i}{2}$ and by defining the linear vibronic couplings $c_i = \frac{\omega_i^2 d_i}{2}$, as well as modifying the reference of the potential energy, the spin-boson Hamiltonian reads

$$\hat{H} = \begin{pmatrix} -\frac{\Delta_{XT-CT}}{2} & V_{XT-CT} \\ V_{XT-CT} & \frac{\Delta_{XT-CT}}{2} \end{pmatrix} + \begin{pmatrix} \frac{1}{2}\sum_{i=1}^{M} p_i^2 + V_+(q_1, \cdots q_M) & 0 \\ 0 & \frac{1}{2}\sum_{i=1}^{M} p_i^2 + V_-(q_1, \cdots q_M) \end{pmatrix} \quad (2)$$

with $V_\pm(q_1, \cdots q_M) = \frac{1}{2}\sum_{i=1}^{M} \omega_i^2 \left(q_i \pm \frac{1}{2}d_i\right)^2$. It has the generic form $H = H_S + H_B + S \cdot B$ with

$$\hat{H}_S = \begin{pmatrix} -\dfrac{\Delta_{XT-CT}}{2} & V_{XT-CT} \\ V_{XT-CT} & \dfrac{\Delta_{XT-CT}}{2} \end{pmatrix}, \quad \hat{H}_B = \dfrac{1}{2}\sum_{i=1}^{M} p_i^2 + \omega_i^2 q_i^2 \quad \text{and} \quad S = \begin{pmatrix} 1 & 0 \\ 0 & -1 \end{pmatrix}, \quad B = \sum_{i=1}^{M} c_i q_i.$$

In order to heuristically take into account a continuous model which is best suited to the polymer material of interest, one can perform a symmetrized Lorentzian broadening of the $M$ modes to obtain a spectral density $J(\omega)$ defined as [56]:

$$\begin{aligned}J(\omega) &= \frac{\pi}{2}\sum_{i=1}^{M} \frac{c_i^2}{\omega_i}\delta(\omega-\omega_i) \\ &\approx \frac{\pi}{2}\sum_{i=1}^{M} \frac{c_i^2}{\omega_i}\left(\frac{\Delta_\omega}{(\omega-\omega_i)^2 + \Delta_\omega^2} - \frac{\Delta_\omega}{(\omega+\omega_i)^2 + \Delta_\omega^2}\right)\end{aligned} \quad (3)$$

where $\Delta_\omega$ is a Lorentzian broadening parameters we take as the root-mean square of the $\omega_i$ frequency spacing according to $\Delta_\omega = \sqrt{\dfrac{1}{M}\sum_{i=1}^{M-1}(\omega_{i+1}-\omega_i)^2}$. This spectral density is assumed to remain the same for all the distances $R$ considered. For the methodology that will follow, i.e. in order to use the HEOM formalism, this spectral density $J(\omega)$ must be parametrized so that the correlation function takes the form of a sum of exponential terms [60]. The choice of the shape of the spectral density and its physical relevance are still widely discussed in the literature for sub-Ohmic [61], Ohmic and super-Ohmic cases [62]. We have chosen to deal with the traditional case of an electron coupled to a phonon bath in solid phase which is known to be super-Ohmic [4]. This choice seems in agreement with the problem of the organic material at stake. The spectral density $J(\omega)$ has been fitted here by $J_0(\omega)$, a set of four-pole Lorentzian functions ensuring a super-Ohmic behavior for small frequencies:

$$J_0(\omega) = \sum_{l=1}^{n_l} \frac{p_l \omega^3}{\left[(\omega+\Omega_{l,1})^2 + \Gamma_{l,1}^2\right]\left[(\omega-\Omega_{l,1})^2 + \Gamma_{l,1}^2\right]\left[(\omega+\Omega_{l,2})^2 + \Gamma_{l,2}^2\right]\left[(\omega-\Omega_{l,2})^2 + \Gamma_{l,2}^2\right]}. \quad (4)$$

Another reason for this choice that differs from the usual Meier-Tannor parametrization [9] is linked to the discussion on the subsequent collective mode model. The integrals defining the effective mode frequency and overall coupling do not converge for the Ohmic case which is not the case for this particular super-Ohmic parameterization.

Four Lorentzian functions have been used and are displayed in Fig. 2. The parameters have been obtained by a non-linear fit algorithm from Gnuplot [63]. They are given in the Appendix.

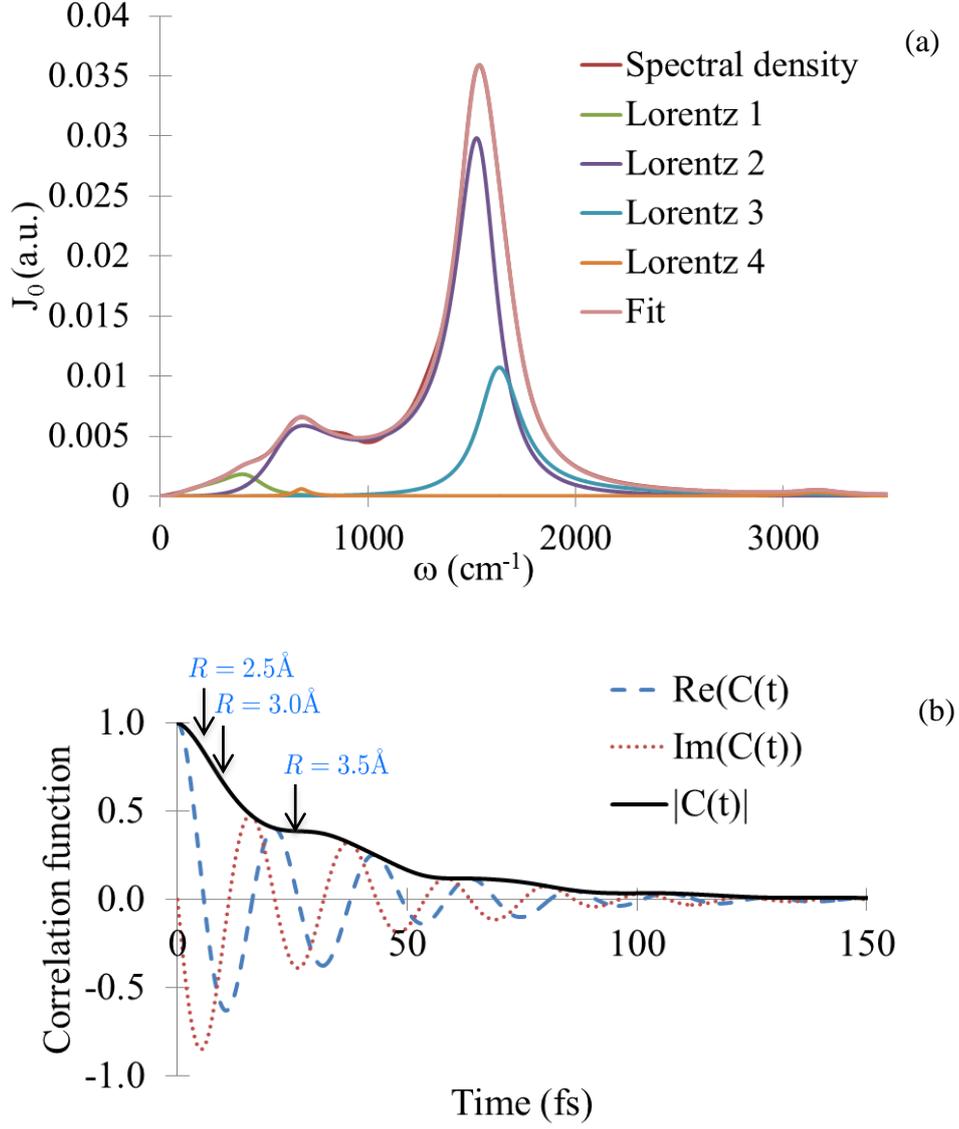

FIG.2. Top: Spectral density $J_0(\omega)$ of the spin-boson model and the fit with four four-pole Lorentzian functions. Bottom: Normalized correlation function corresponding to the spectral density shown in the upper panel. The arrows indicate the Rabi periods for the three inter fragment distances.

The basic tool to discuss memory effects is the bath correlation function defined by

$$C(t-\tau) = Tr_B\left[B(t)B(\tau)\rho_B^{eq}\right] \qquad (5)$$

where $B(t) = \exp(i\hat{H}_B t)\hat{B}\exp(-i\hat{H}_B t)$ is the bath operator in the Heisenberg representation and $\rho_B^{eq} = \exp(-\beta\hat{H}_B)/Tr_B\left[\exp(-\beta\hat{H}_B)\right]$ is the Boltzmann equilibrium density matrix of the bath. In virtue of the fluctuation-dissipation theorem, one may write :

$$C(t-\tau) = \frac{1}{\pi} \int_{-\infty}^{+\infty} d\omega \frac{J(\omega) e^{i\omega(t-\tau)}}{e^{\beta\omega} - 1}. \quad (6)$$

where $\beta = 1/k_B T$. The correlation function is shown in Fig. 2. The comparison between the correlation time and the Rabi periods is the only *a priori* tool to predict a non Markovian behavior. Here all the Rabi periods are smaller (for $R$ = 2.5 and 3 Å) or close to the typical decay time of the correlation function (for $R$ = 3.5 Å) so that the $R$ = 2.5 Å case is expected to exhibit the most non-Markovian dynamics for the reduced system.

As shown in Sec. IV, a non-Markovianity signature can be obtained using the exact dynamical map, i.e. propagations of basis operators are required and they are done here with the HEOM method, presented in the next section.

### III. HIERARCHICAL EQUATIONS OF MOTION

The hierarchical equations of motion method (HEOM) [8] is a powerful tool to solve the master equation with a linear coupling to a harmonic oscillator bath. It takes advantage of the fact that the previous approximations (linear coupling and harmonic oscillators) define a Gaussian bath which means that the second order in the cumulant expansion will be exact to accurately describe the time evolution of the reduced density matrix. Defining the reduced density matrix corresponding to the system, in interaction representation by $\rho_I(t) = e^{i\hat{H}_S t} \rho(t) e^{-i\hat{H}_S t}$ (with $\hbar = 1$), its time evolution is given by

$$\rho_I(t) = Tr_B \left[ e^{\int_0^t d\tau L(\tau)} \rho_B^{eq} \rho_I(0) \right] = e^{\int_0^t d\tau \int_0^\tau dt' Tr_B [L(\tau) L(t') \rho_B^{eq}]} \rho_I(0) \quad (7)$$

where $L(t) \bullet = -\frac{i}{\hbar} [S(t) B(t), \bullet]$ is the Liouvillian of the system-bath interaction with the system coupling operator in interaction representation $S(t) = e^{i\hat{H}_S t} \hat{S} e^{-i\hat{H}_S t}$ and the bath operator as given above.

In order to obtain a computationally efficient algorithm, one has to express the bath correlation function as a sum of complex exponential functions :

$$C(t-\tau) = \sum_{k=1}^{n_{cor}} \alpha_k e^{i\gamma_k (t-\tau)} \quad (8)$$

This can be achieved by using the parametrization of the spectral density of equation (4), where the integration in Eq. (6) is performed analytically yielding explicit expressions for $\alpha_k, \gamma_k$ (see Appendix). One can also express the complex conjugate of the correlation function by keeping the same coefficients $\gamma_k$ in the exponential functions with modified coefficients $\tilde{\alpha}_k$ according to:

$$C^*(t-\tau) = \sum_{k=1}^{n_{cor}} \tilde{\alpha}_k e^{i\gamma_k (t-\tau)} \quad (9)$$

with $\tilde{\alpha}_{l,1} = \alpha_{l,2}^*$, $\tilde{\alpha}_{l,2} = \alpha_{l,1}^*$, $\tilde{\alpha}_{l,3} = \alpha_{l,4}^*$, $\tilde{\alpha}_{l,4} = \alpha_{l,3}^*$ and $\tilde{\alpha}_{j,matsu} = \alpha_{j,matsu}$ where the $\alpha_{l,m}$ with $m = 1, 4$ are related to the four poles of each Lorentzian $l$ (see Appendix for details).

With these expressions for $C(t-\tau)$ and $C^*(t-\tau)$, the master equation in interaction representation can be written as a time-local hierarchical system of coupled differential equations (each matrix can communicate only with the superior and inferior level in the hierarchy) :

$$\dot{\rho}_{\mathbf{n}}(t) = i\sum_{k=1}^{n_{cor}} n_k \gamma_k \rho_{\mathbf{n}}(t) - i\left[S(t), \sum_{k=1}^{n_{cor}} \rho_{\mathbf{n}_k^+}(t)\right] - i\sum_{k=1}^{n_{cor}} n_k \left(\alpha_k S(t) \rho_{\mathbf{n}_k^-} - \tilde{\alpha}_k \rho_{\mathbf{n}_k^-} S(t)\right) \quad (10)$$

with the collective indices $\mathbf{n} = \{n_1, \cdots, n_{n_{cor}}\}$, $\mathbf{n}_k^+ = \{n_1, \cdots, n_k + 1, \ldots, n_{n_{cor}}\}$, and $\mathbf{n}_k^- = \{n_1, \cdots, n_k - 1, \ldots, n_{n_{cor}}\}$. In this hierarchy of auxiliary density matrices, the system density matrix is given by top row, i.e. for $\mathbf{n} = \{0, \cdots, 0\}$ hence $\rho_I(t) = \rho_{\{0,\cdots,0\}}(t)$. The level of the hierarchy is chosen until convergence is reached for the system density matrix.

Taking into account the full time dependence of $C(t)$, via Eqs. (6), (8) and (9), implies that the dynamical modelling based on the HEOM methodology takes the full frequency dependence of the spectral density into account. This aspect is very important, in particular when analyzing non-Markovian aspects, which mainly stem from the low frequency part of $J(\omega)$ which are not in resonance with any system transition.

## IV WITNESS OF NON-MARKOVIANITY

Different measures have been proposed to quantify the non-Markovianity [31, 32, 33]. Breuer's non-Markovianity measure [35, 64, 65] is based on the trace distance of pairs of initial states $D(\rho_1 - \rho_2) = 1/2 \|\rho_1 - \rho_2\|$. In a Markovian evolution, this trace distance monotonously decreases leading to indistinguishability of the quantum states while any transitory increase is a signature of a back flow from the bath. This measure has been used recently to quantify non-Markovianity in photosynthetic complex [66]. However, this measure requires a sampling of pairs of initial states. Other measures based on the volume of dynamically accessible states in the system [36] or the time-dependent decoherence canonical rates [31] require propagation of basis operators only.

In the two-level case ($d = 2$), the dynamical map $\rho(t) = \phi_t[\rho(0)]$ can be decomposed in the basis set of $d^2$ Hermitian operators formed by the identity $G_0 = \mathbf{I}/\sqrt{d}$ and three operators $G_m$ with $m = 1, 3$ which are the three Pauli matrices $\boldsymbol{\sigma}_{\mathbf{x,y,z}}/\sqrt{d}$. The generalization for higher dimension is given in ref. [36]. The dynamical map is then written

$$\rho(t) = \sum_{k=0}^{d^2-1} Tr(G_k \rho(0)) \phi_t[G_k]. \quad (11)$$

The volume of accessible states $V(t) = \det(\mathbf{F})$ can be computed as the determinant of the dynamical map matrix in this basis set

$$F_{m,n}(t) = Tr(G_m \phi_t[G_n]). \quad (12)$$

Any increase of this volume is considered to be a signature of a non-Markovian behavior. However, this measure could fail to detect non-Markovianity because it depends on the sum of the canonical rates that appear in the canonical form of the master equation (see below).[31] The sum may be positive even if some individual rate is negative.

The time non-local master equation can always be recast in a canonical Lindblad with time dependent rates associated with decoherence decay channels. Details and demonstration can be found in ref. [31]. First, the master equation is reformulated as

$$\dot{\rho}(t) = -\frac{i}{\hbar}\left[\hat{H}_S, \rho(t)\right] + \sum_k A_k(t)\rho(t)B_k(t) \tag{13}$$

where $A_k$ and $B_k$ are system operators which can be expressed in terms of the complete set of the Hermitian operators $G_m$

$$\dot{\rho}(t) = -\frac{i}{\hbar}\left[\hat{H}_S, \rho(t)\right] + \sum_{j,k=0}^{d^2-1} c_{jk}(t)G_j\rho(t)G_k. \tag{14}$$

By separating terms related to the $G_0$ operator, i.e. terms containing coefficients $c_{j0}$, this master equation is rewritten with a corrected system Hamiltonian $\hat{H}_{Scor}$ including these terms [31] and a relaxation operator involving only the operator associated with the Pauli matrices

$$\dot{\rho}(t) = -\frac{i}{\hbar}\left[\hat{H}_{Scor}, \rho(t)\right] + \sum_{j,k=1}^{d^2-1} D_{jk}(t)\left(G_j\rho(t)G_k - \frac{1}{2}\{G_kG_j, \rho(t)\}\right). \tag{15}$$

The key is the eigenvalues and eigenvectors of the Hermitian decoherence matrix $D_{jk}(t)$ which provide the canonical Lindblad form with time dependent canonical decoherence rates $g_k(t)$ and the canonical decoherence channels $C_k(t)$

$$\dot{\rho}(t) = -\frac{i}{\hbar}\left[\hat{H}_{Scor}, \rho(t)\right] + \sum_{k=1}^{d^2-1} g_k(t)\left(2C_k(t)\rho(t)C_k^\dagger(t) - \{C_k^\dagger(t)C_k(t), \rho(t)\}\right) \tag{16}$$

where $D_{ij}(t) = \sum_{k=1}^{d^2-1} U_{ik}(t)g_k(t)U_{jk}^*(t)$ and $C_k(t) = \sum_{i=1}^{d^2-1} U_{ik}(t)G_i$.

The occurrence of negative canonical decoherence rates $g_k(t)$ yields an alternative characterization of non-Markovianity. Moreover, the rates are directly linked to the time evolution of the volume of accessible states through the relation $V(t) = V(0)\exp\left(-d\int_0^t \Gamma(s)ds\right)$ with $\Gamma(t) = \sum_{k=1}^{d^2-1} g_k(t)$ and $d$ is the dimensionality of the reduced system (here, $d=2$). Hence, one sees that the criteria given by the volume can be considered as an average measure, based on the sum of the rates which can remain positive even if one of them is negative. This will be clearly evidenced in the Results section.

A possible numerical strategy to compute the decoherence matrix $D_{ij}(t)$ giving access to the rates $g_k(t)$ and thus to the volume $V(t)$ is given in the Appendix.

## V. RESULTS OF THE SB MODEL

The electron transfer dynamics is treated by HEOM at level 6 of the hierarchy (i.e. order 12 in a perturbative expansion). These orders were found to give converged results. The requirement of such high orders for converged results clearly shows that the regime is strongly non perturbative.

A first criterion in order to discuss non-Markovianity is the ratio between the bath correlation time and the characteristic timescale of the system dynamics, here the Rabi period of the electronic system. As seen in Fig.2, the typical decay time of the bath correlation time is in the order of $\tau_{cor} \approx 30\,\text{fs}$. The Rabi periods for the different internuclear distances, given in Table 1, are smaller than this decay time $\tau_{cor}$ and consequently, a non-Markovian behavior is expected for all of them, but most pronounced for the case $R = 2.5$ Å. Another indicator is the position of the Rabi frequency (see Table 1) with respect to the maxima of the spectral density. As illustrated by Breuer in a SB model with an Ohmic spectral density with a Lorentzian cutoff, a more Markovian behavior is expected when the transition frequency corresponds to the maximum of $J(\omega)/(e^{\beta\omega}-1)$ which at room temperature is close to the value of the spectral density at this given frequency [67]. In the case $R = 2.5$ Å, the Rabi frequency is off-resonance with all the structured bands of the spectral density so that a more non-Markovian behavior could be expected from both criteria.

### A. Dynamics of the system

As a test case, the initial state is prepared with an electronic population in the diabatic XT state and the reference coordinates of the bath modes are chosen as zero in agreement with Eq. (2). The time dependent population in the XT state is displayed in Figure 3a. The asymptotic diabatic populations are reached in about 200 fs for the case $R = 3$ Å and $R = 3.5$ Å while it takes 2 ps in the $R = 2.5$ Å case. The difference of behavior will be qualitatively discussed from the effective potential energy curves in section VIa. The purity of the system measured by $Tr[\rho^2(t)]$ is given in Figure 3b. It illustrates the fast decoherence and the evolution towards the final statistical mixture at thermal equilibrium. In the cases $R = 2.5$ Å and $R = 3$ Å, the final purity is nearly equal to the Boltzmann distribution of the adiabatic electronic state. The mixture is the practically a single state so one observe an increase of the purity towards nearly 1. In the third case $R = 3.5$ Å, the final value different from one can be taken as a sign of the strong correlation between the system and the bath resulting in a noncanonical distribution in the system [68]. Figures 3c and 3d also show the early evolution of the average energy $E = Tr[\rho H_S]$ and of the von Neumann entropy $S = -Tr(\rho \log_2 \rho)$. The relaxation is not completely monotonous exhibiting very weak energy back flows towards the system. The entropy that is closely linked to the purity, also presents a non-monotonous early evolution. When the purity has a local bump, the system entropy decreases and this has been propose as a signature of non-Markovianity since the entropy decrease can be interpreted as an information back flow from the environment [69]. This aspect will be further discussed in relation to the non-Markovianity measures to be presented in the following sections.

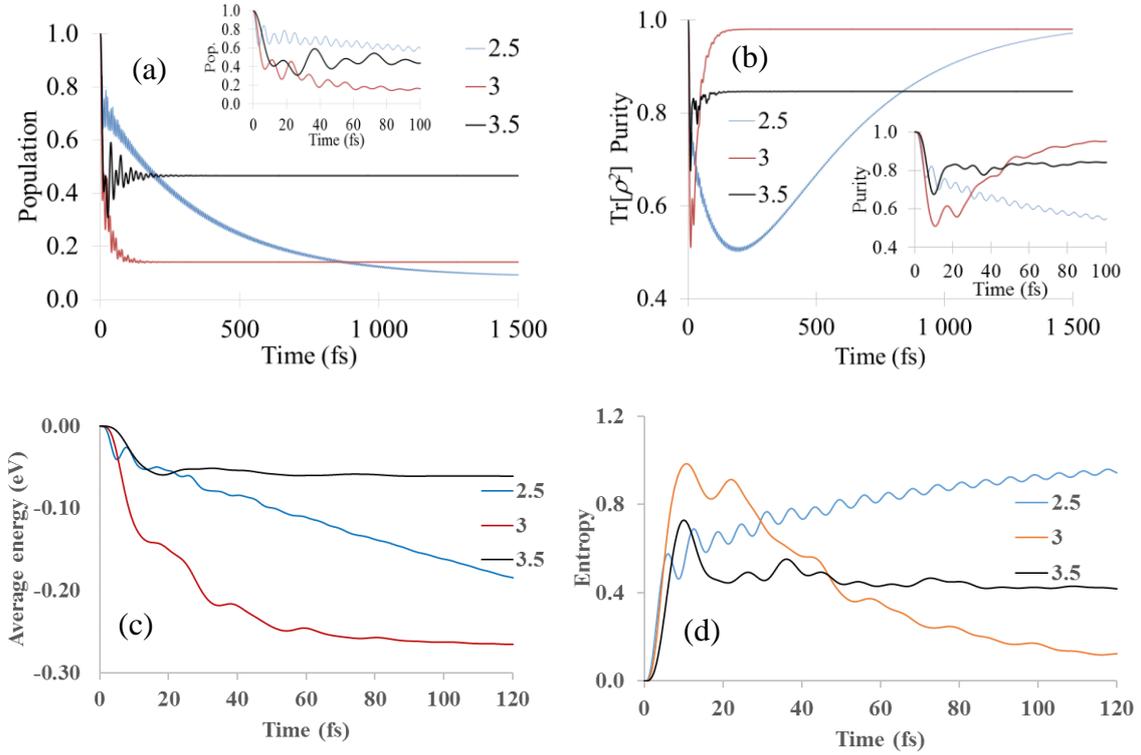

FIG. 3 Dissipative dynamics of the electronic system prepared in the XT state for three inter-fragment distances in Å. Panel a : evolution of the population in the XT state; panel b : purity; panel c : average energy; panel d : von Neumann entropy.

**B. Canonical rates**

In order to study the non-Markovian behavior, dynamics is carried out not only with a single initial state but with each of the $G_m$ matrices ($m$ = 1,4) as defined in Eq. (11) and following the procedure defined in Appendix. The three canonical rates $g_k(t)$ and their sum $\Gamma(t)$ are displayed in Fig. 4. Several observations arise. In the three cases, one negative rate exists and the sum transitorily becomes negative. In the $R$ = 3.5 Å case, some narrow negative and positive peaks appear leading to a compensation in the sum. However, as seen in the inset there is also a negative total rate of the same order of magnitude than that obtained in the two other examples, for instance in the range 20-30 fs. It remains a debated question whether the occurrence of a negative rate is sufficient to induce a non-Markovian behavior. We may wonder about the signification of negative peaks compensated by positive ones. It seems that a transitory negative sum $\Gamma(t)$ is a more significant criterion. Another argument is that for other examples treated in the Markovian Redfield approach without secular approximation, the diagonalization of the decoherence matrix gives constant rates with one negative leading however to a positive sum. Hence, it seems difficult to use the behavior of the individual rates to quantify the non-Markovianity. Indeed, the large negative peaks in the $R$ = 3.5 Å case could lead to a conclusion opposite to the criterion based on the comparison of the time scale. We will see below that the measure of the volume of reachable states is in better agreement with the expected degree of non-Markovianity based on timescale arguments. On the other hand, all the rates present an oscillatory pattern corresponding to a typical period of about 20 fs which we will try to link with the motion of a collective bath mode analyzed from the HEOM auxiliary matrices. (see section V.D and section VI.).

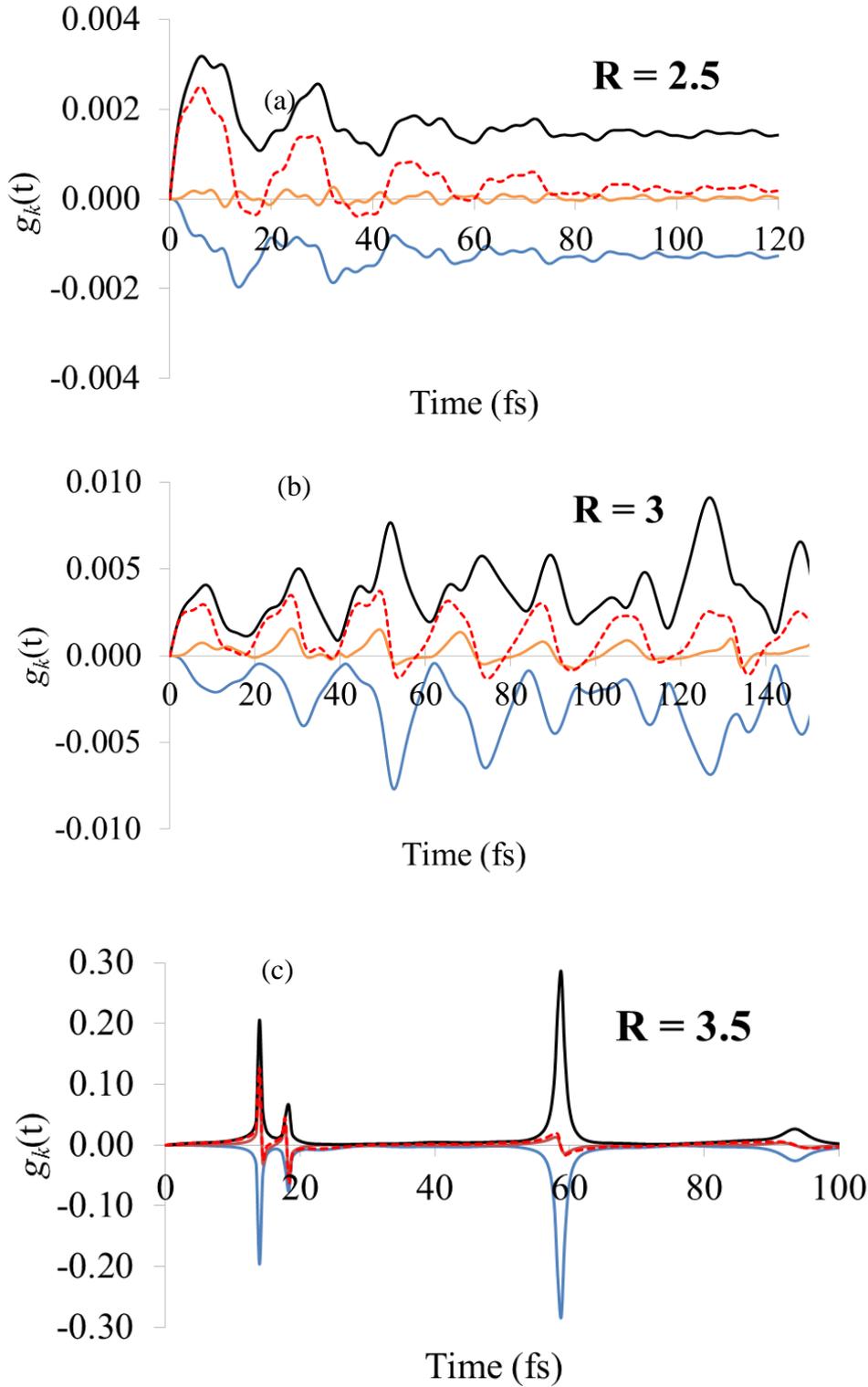

FIG.4 Canonical decoherence rates $g_k$ (Eq. (16)) (full lines) and their sum (dashed line) of the SB model for the three inter-fragment distances in Å.

## C. Volume of accessible states

The volume of the accessible states in the Bloch sphere is presented in Fig. 5. The decrease of the volume is always ultrafast and some non-Markovianity witness can be observed.

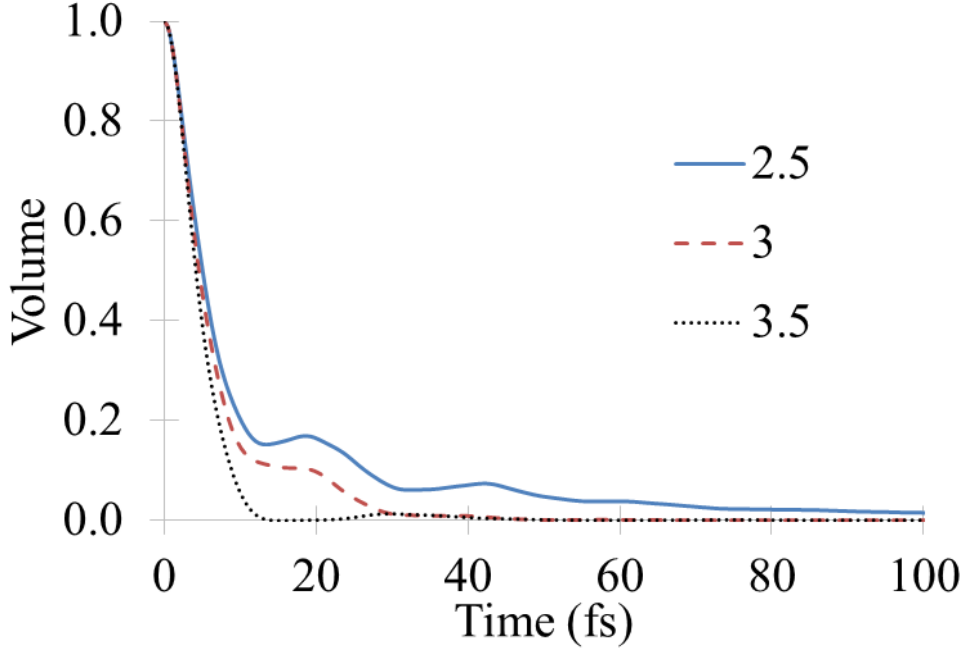

FIG. 5 Volume of the accessible states in the Bloch sphere of the SB model for three inter-fragment distances in Å.

In the case $R = 2.5$ Å, the dynamics is clearly non Markovian since the volume $V(t)$ presents two bumps around 20 and 40 fs corresponding to the two time intervals where the sum of the rates becomes negative. The bumps are less pronounced in the $R = 3.0$ Å case but the profile remains far from a Markovian exponential decay. This is in agreement with the behavior of the sum of the rates that presents very weak negative values in this timescale. For $R = 3.5$ Å, the volume has a more Markovian profile with a very weak bump around 30 fs which can be correlated to the negative sum in this interval 20-30 fs. The signature of some non-Markovianity in the volume interestingly corresponds to the expected behavior from the comparison of the Rabi periods with the correlation time.

The non-monotonic evolution of the volume of accessible states might be connected to similar properties of the dynamical map and therefore to non-monotonic evolution of some properties such the average energy or the entropy. This effect has already been observed in Fig. 3. For instance, the slow-down of the energy decay around 20fs, in the case $R = 3$ Å approximately corresponds to the plateau in the evolution of the volume (dashed line in Fig. 5). It is also the time where the entropy has a dip. However, the correlation is not as clear for the other cases. The very early increase in the energy, in the case $R = 2.5$ Å is not correlated to any bump in the volume. The volume of accessible state is a global property of the map and depends on the sum of the canonical decoherence rates. On the contrary, $E$ or $S$ depends on the initial state. Each state has a different time-dependent decomposition on the decoherence channels and the rates have not the same weights. This opens questions about the interest of controlling non-Markovianity to enhance or remove the back flows from the surrounding and therefore the effects on $E$ or $S$ by interaction with fields of by engineering the surrounding.

Now, we shall link these transitory weak oscillations with the dynamics of the collective mode of the bath obtained from the HEOM method.

**D. Signature of the bath dynamics**

A strategy to get insight into the bath dynamics from HEOM has been proposed by Shi *et al.*[70] These authors showed that the moments $X^{(n)}(t) = Tr_B\left[B^n \rho_{tot}(t)\right]$ of the collective mode $B = \sum_i c_i q_i$ can be obtained by the HEOM auxiliary matrices. In particular, the expectation value of $B$ is given by the sum of the first level auxiliary matrices $X^{(1)}(t) = -\sum_{\tilde{\mathbf{n}}} \rho_{\tilde{\mathbf{n}}}(t)$, where the sum runs over all index vectors $\tilde{\mathbf{n}} = \{n_1, \cdots, n_{n_{cor}}\}$ with $\sum_l n_l = 1$ (see Eq.(12) of reference 70). This quantity provides a signature of the induced bath dynamics. As discussed in Eq. (18) of reference 70, one can recast the master equation to emphasize the role of the $X^{(1)}(t)$ matrix in the system dynamics by writing

$$\dot{\rho}(t) = -i\left[\hat{H}_S, \rho(t)\right] + i\left[S, X^{(1)}(t)\right]. \qquad (17)$$

The same initial state as in Section V.A. has been used. Figure 7 displays the diagonal elements of $X^{(1)}(t)$ for the two XT and CT electronic states. They provide information about the damped oscillation of the expectation value of this bath collective mode towards the equilibrium asymptotic state in each electronic state. One can observe that the electron transfer induces a strong reorganization of the bath. A basic period of about 20 fs appears and this nicely corresponds to the period seen in the real and imaginary part of the correlation function (21fs) and the one observed in the evolution of the sum of the canonical rates or in the volume of accessible states (see Figs 4 and 5). The period of 21.4 fs also corresponds to that predicted by the maximum of the highest peak in the spectral density at 1550 cm$^{-1}$. The early motion in the XT state reveals the high system-bath correlation mainly in the $R = 2.5$ and 3 cases. The first part of the vibration is modulated to the electronic Rabi period (6.3 and 12.3 fs respectively) while the following regime exhibits the nearly 20 fs period, more precisely 21, 20 and 19 fs respectively). On the other hand, the survival of damped oscillations in the XT or CT states is also a signature for the electronic transfer timescale. In the $R = 2.5$ Å case, one sees the slow evolution towards the equilibrium final state already observed in the purity evolution (see Fig. 3). For the $R = 3$ Å case, the transition is fast and the damping mainly occurs in the second electronic state while in the last example, for $R = 3.5$ Å, the damping is similar in the two electronic states which remain equally populated. In every case, the fundamental frequency is about 18 fs.

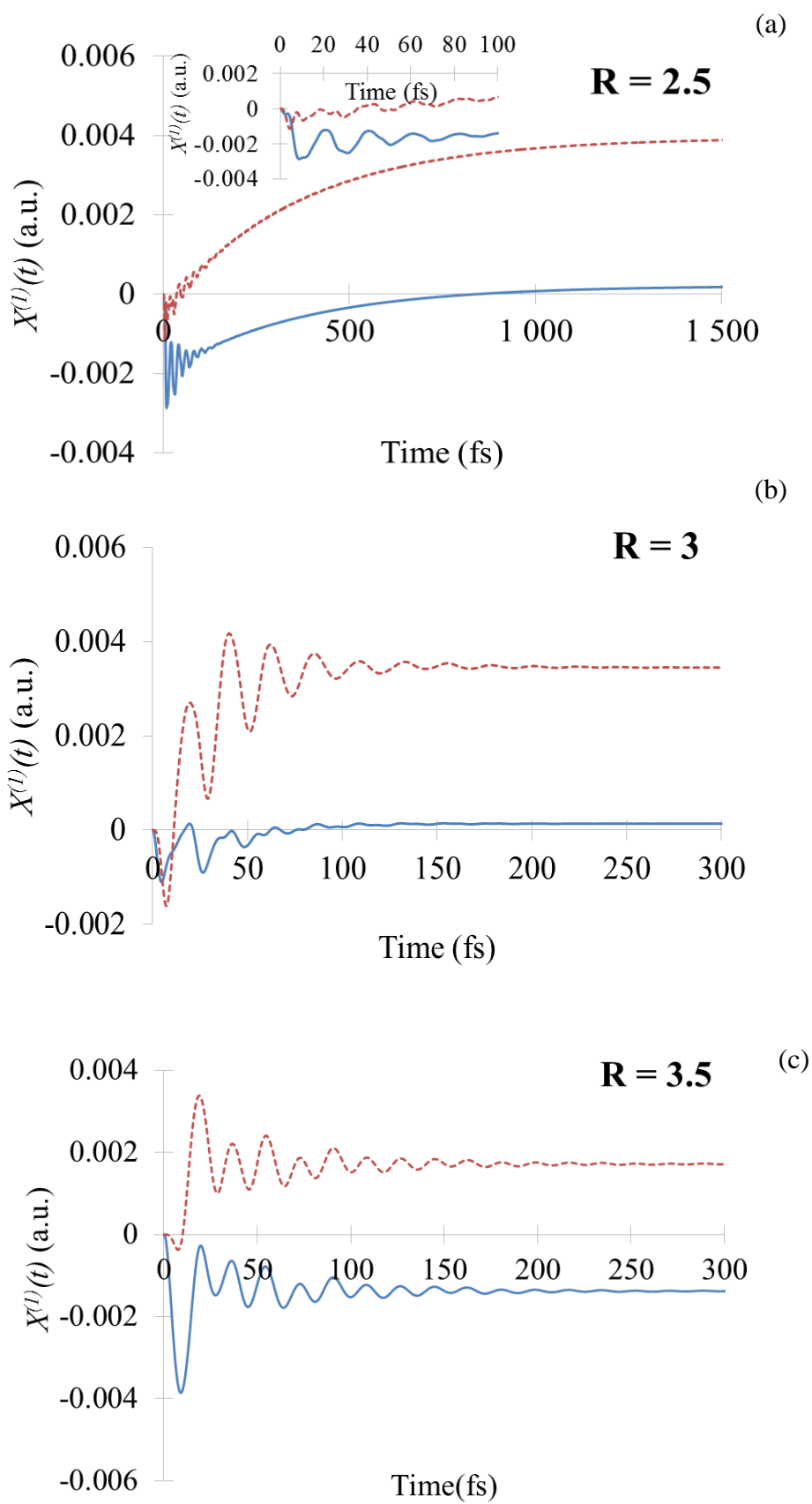

FIG. 6 Diagonal elements of the $X^{(1)}(t)$ operator giving the first moment of the bath collective mode $B = \sum_i c_i q_i$ for three inter fragment distances in Å. Full line : XT state, dashed line : CT state.

The information about the bath dynamics provided by the auxiliary matrices of HEOM reveals an order of magnitude for the frequency of the effective vibrational mode that is coupled to the electron transfer. However, it is not sufficient to build a vibronic one-dimensional model that requires the displacement of the oscillators. Hence in the next section, we compare these findings with results obtained from the effective mode theory already used in a previous paper [53].

## VI VIBRONIC SB MODEL

Constructing a vibronic SB model presents many interesting features that we will detail in this section. First of all, we have seen in the previous section that the bath collective mode nicely reproduces the evolution period of the sum of the canonical rates. In order to understand this result, one first tries to extract the parameters of this mode from the effective mode strategy [48, 49, 50, 52] and its associated correlation function. We carry out a systematic analysis to check the validity of the Markovian approximation to treat the effective vibronic system. Then, if the movement of the effective mode can drive most of the non-Markovian behavior, it could constitute an excellent guess for instance to control the information flow back to the system and limit as well the decoherence process even in complex molecular systems.

### A. **Effective mode**

We carry out a change of coordinates to define a primary collective mode that will be included in the system providing a coordinate representation of the SB model [1, 17]. This mapping towards a vibronic or reaction path model has been frequently used in the literature [1, 71, 72, 73]. The collective mode, often called 'effective mode' can still be coupled to a secondary bath. Different definitions of a collective mode have been proposed [49, 50, 52, 74, 75]. We have chosen here the direction which endorses all the vibronic couplings, i.e. the only direction inducing a variation of the electronic energy gap.

Defining a canonical basis $\mathbf{e}_i, (i=1\cdots M)$ to express the normal modes as a vector according to $\mathbf{q} = \sum_{i=1}^{M} q_i \mathbf{e}_i$ and the displacements according to $\mathbf{c} = \sum_{i=1}^{M} c_i \mathbf{e}_i$, we aim at constructing a new basis $\mathbf{g}_i, (i=1\cdots M)$ within this vector space, such that one of the new basis vectors, denoted $\mathbf{g}_1$, comprises the coupling to the bath. Consequently, we choose

$$\mathbf{g}_1 = \frac{\mathbf{c}}{\|\mathbf{c}\|} = \frac{\mathbf{c}}{D_0} \qquad (18)$$

with $D_0 = \|\mathbf{c}\| = \sqrt{\sum_{i=1}^{M} c_i^2}$. The remaining new basis vectors, $\mathbf{g}_i, (i=2,\cdots M)$ are constructed to be normalized and orthogonal to $\mathbf{g}_1$, as well as orthogonal within their own space [76]. Since the potential energy operator appearing in Eq. (2) can be written in terms of $\mathbf{q}$ and $\mathbf{c}$ as

$$V^{\pm}(\mathbf{q}) = \frac{1}{2}\sum_{i=1}^{M} \omega_i^2 q_i^2 \pm \sum_{i=1}^{M} c_i q_i = \frac{1}{2}\mathbf{q}^T \boldsymbol{\omega}^2 \mathbf{q} \pm \mathbf{c}\mathbf{q} \qquad (19)$$

we can express **q** in terms of the new basis $\mathbf{g}_i$ according $\mathbf{q} = x_1\mathbf{g}_1 + \sum_{i=2}^{M} x_i\mathbf{g}_i$ leading to the potential energy functions $V^\pm$ expressed in the new coordinates $x_i$:

$$V^\pm(\mathbf{x}) = \frac{1}{2}\left[\Omega_1^2 x_1^2 + \sum_{i=2}^{M}\tilde{\omega}_i^2 x_i^2 + \sum_{i=2}^{M}\kappa_i x_i x_1\right] \pm D_0 x_1 \qquad (20)$$

where $\Omega_1^2 = \frac{1}{D_0^2}\sum_{i=1}^{M} c_i^2 \omega_i^2$ is the square of the effective mode frequency, $\tilde{\omega}_i^2$ are the square of the secondary mode frequencies and $\kappa_i$ the vibronic coupling of the secondary modes to the effective coordinate [76].

The potential energy operator exhibits the desired features: one harmonic oscillator is extracted from the bath and $M-1$ other oscillators are coupled to this one. The same transformation has to be applied to the kinetic operator as well. But taking into account that it is a linear transformation of the coordinates and the $\{\mathbf{g}_i\}$ is a set of orthonormal vectors, the kinetic operator merely leads to $p_i^2 = \tilde{p}_i^2$. Hence, the transformed Hamiltonian can be expressed as

$$\hat{H} = \hat{H}_{vib} + \begin{pmatrix} x_1 & 0 \\ 0 & x_1 \end{pmatrix}\sum_{i=2}^{M}\kappa_i x_i + \hat{H}_b^{(sec)} \qquad (21)$$

with the vibronic Hamiltonian $\hat{H}_{vib}$ and the secondary bath $\hat{H}_b^{(sec)}$ given by

$$\hat{H}_{vib} = \begin{pmatrix} \frac{1}{2}p_1^2 + \frac{1}{2}\Omega_1^2 x_1^2 - D_0 x_1 - \frac{\Delta_{XT-CT}}{2} & V_{XT-CT} \\ V_{XT-CT} & \frac{1}{2}p_1^2 + \frac{1}{2}\Omega_1^2 x_1^2 + D_0 x_1 + \frac{\Delta_{XT-CT}}{2} \end{pmatrix}$$

and $\hat{H}_b^{(sec)} = \frac{1}{2}\sum_{i=2}^{M}\left(p_i^2 + \tilde{\omega}_i^2 x_i^2\right)$ respectively. This Hamiltonian has the generic form of two shifted harmonic oscillators along a coordinate $x_1$ (see Fig.7) which is coupled to a secondary bath of harmonic oscillators. In the continuous limit, the effective mode frequency $\Omega_1$ and the renormalization coupling constant $D_0$ can be expressed as [75]:

$$D_0^2 = \frac{2}{\pi}\int_0^{+\infty} d\omega J_0(\omega)\omega \qquad (22)$$

$$\Omega_1^2 = \frac{2}{\pi D_0^2}\int_0^{+\infty} d\omega J_0(\omega)\omega^3. \qquad (23)$$

It is noteworthy to stress that those two integrals are analytical for the super-Ohmic case in the present paper. However, it is not the case for the Ohmic one where the integral giving $\Omega_1$ diverges and a cutoff frequency is required [53].

In order to calculate the $\kappa_i$, one can discretize the spectral density in a sample of equidistant frequencies and use projection techniques [76] or directly use a continuous limit defining a secondary spectral density with the following formula [52, 75]:

$$J_1(\omega) = \frac{D_0^2 J_0(\omega)}{J_0^2(\omega) + W_{0,PV}^2(\omega)} \qquad (24)$$

$$W_{0,PV}(\omega) = \frac{1}{\pi} PV \int_{-\infty}^{+\infty} d\omega' \frac{J_0(\omega')}{\omega' - \omega} \qquad (25)$$

where $W_{0,PV}(\omega)$ is the principal value of Hilbert transform of the primary spectral density.

Analyzing the vibronic Hamiltonian already gives a hint towards the expected dynamical behavior : The values of the parameters of the primary effective mode are $\Omega_1$ = 2084 cm$^{-1}$ and $D_0$ = 5.894.10$^{-4}$ a.u. This effective mode is blue-shifted from the maximum of the spectral density. The corresponding period is 16 fs. It is close but slightly shorter than the vibrational period detected by the HEOM auxiliary matrices. However, the vibrational wave packet does not feel a strict harmonic potential due to the non-adiabatic interaction. We then estimate the characteristic timescale from the average of the vibronic transitions weighted by the matrix elements of the coordinate

$$\tau_{EM} = \frac{2\pi}{\Omega_{EM}} \text{ with } \Omega_{EM} = \frac{\sum_{ij} \Delta E_{ij}(x_1)_{ij}}{\sum_{ij}(x_1)_{ij}} \qquad (26)$$

The choice of the basis size of the coordinate ensures the conservation of the trace at less than 1%. By this way, the effective characteristic timescale of the average position is also around 19 fs (18.4, 18.2 and 20.2 fs respectively). As the HEOM bath mode dynamics after the early transitory regime is mainly a damping with nearly the same period, the effective mode model is expected to provide relevant qualitative results in this example if the residual bath is Markovian and the second order sufficient as will be discussed below.

By using the diabatic electronic parameters for the three inter fragment distances (see Table 1 and Fig. 7), the vibronic model evolves from an inverse Marcus donor-acceptor shape for $R$ = 2.5 and 3Å to a normal Marcus profile for $R$ = 3.5Å [53]. This agrees with the behavior of the population evolution seen in Fig. 3. The very low relaxation in the $R$ = 2.5Å case can be understood by a trapping of the vibrational wave packet below the crossing point in an inverse Marcus situation. On the contrary, in the case $R$ = 3Å the crossing is at the minimum of the XT state leading to a very fast depopulation and the case $R$ = 3.5Å corresponds to a nearly degenerate double well so a fifty-fifty equilibrium population is expected.

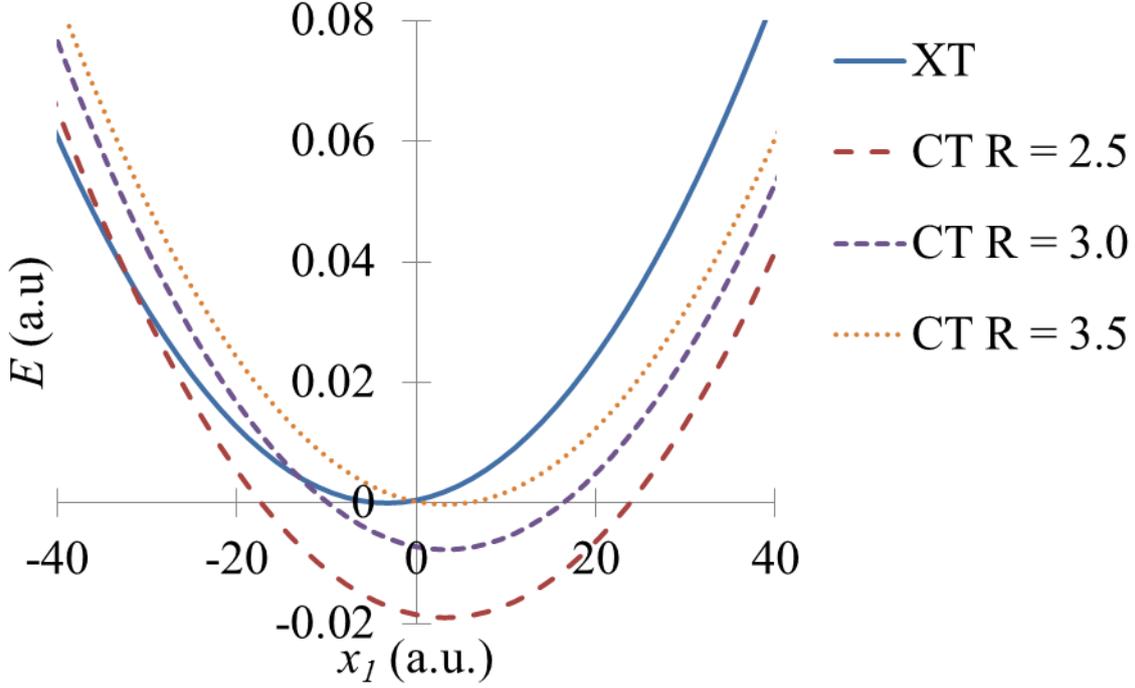

FIG. 7 Effective mode representation of the spin-boson model. XT and CT diabatic curves are represented for different intermolecular distances $R$.

Based on Eq. (21), we thus treat the new extended system coupled to the secondary bath within standard Redfield theory, i.e. within 2$^{nd}$ order perturbation theory and Markov approximation. Treating this extended system coupled to the secondary bath also within HEOM is currently numerically inaccessible. In order to validate the more approximate approach, we thoroughly check both the Markov approximation and the perturbative character of the coupling to the secondary bath. As will be shown below, both approximations are valid in our case, thus allowing to treat the ET dynamics after including the effective mode into an extended system within the much simpler Redfield approach.

To numerically treat the vibronic model including the harmonic effective coordinate $x_1$, it is represented in an eigenbasis of $\hat{H}_{vib}$ comprising $N$ states, which is obtained by diagonalizing the Fourier-grid Hamiltonian [77] constructed within the space of diabatic representation. Hence the vibronic eigenfunctions can be expressed as $|\psi_\nu\rangle = |\chi_\nu^{(CT)}\rangle|CT\rangle + |\chi_\nu^{(XT)}\rangle|XT\rangle$ where $|\chi_\nu^{(CT),(XT)}\rangle$ are wavefunctions of diabatic basis with $\nu = 1, 2, \ldots, N$.

The first issue one can encounter when dealing with the collective coordinate is the definition of the initial state in a comparable way with the spin-boson. As throughout all our calculations, we assume a factorizable system-bath initial condition in order to mimic the initial state of the calculations presented in Section V. A. The initial total density matrix can then be written as

$$\rho_{tot}(0) = |XT\rangle\langle XT| \otimes \rho_{B,eq} = |XT\rangle\langle XT| \otimes \frac{e^{-\beta H_B}}{Tr_B\left[e^{-\beta H_B}\right]} \tag{27}$$

with

$$\hat{H}_B = \frac{1}{2}\sum_{i=1}^{M}\left(p_i^2 + \omega_i^2 q_i^2\right) = \frac{1}{2}\left(p_1^2 + \Omega_1^2 x_1^2\right) + \hat{H}_b^{(sec)} + \frac{1}{2}x_1\sum_{i=2}^{M}\kappa_i x_i.$$

In the second-order perturbative approximation which will be later be used for the secondary bath modes, we can assume that the coupling to the secondary bath is small, thus, the system density matrix for the effective mode can be expressed as a factorizing density matrix, comprising the Boltzmann distribution of the secondary bath, and a Boltzmann distribution of the vibrational states within the diabatic electronic state $|XT\rangle$:

$$\rho_{tot}(0) \approx \rho_S(0) \otimes \left[\frac{e^{-\beta H_b^{(sec)}}}{Tr_{\tilde{B}}\left[e^{-\beta H_b^{(sec)}}\right]}\right] \quad (28)$$

$$\rho_S(0) \approx |XT\rangle\langle XT|\frac{e^{-\beta H_\Omega}}{Tr_{x_1}\left[e^{-\beta H_\Omega}\right]}$$

where $H_\Omega = \frac{1}{2}\left(\tilde{p}_1^2 + \Omega_1^2 x_1^2\right)$. The corresponding system density matrix can then be projected on the electronic degrees of freedom leading to the reduced electronic density matrix :

$$\rho_{el,ij} = Tr_{x_1}\left[\rho_{S,ij}\right] = \sum_{v,v'}|\chi_v^{(i)}\rangle\langle\chi_{v'}^{(j)}|\langle\psi_v|\rho_S|\psi_{v'}\rangle \quad (29)$$

where $\{i,j\} = \{XT, CT\}$ are the electronic degrees of freedom.

**B. Results of the vibronic model**

As the coordinate $x_1$ representing the collective mode already gathers all the electronic-bath coupling since it is the only displacement inducing a variation of the electronic gap, it can drive most of the electronic dynamics. It is this part we want to assess.

The collective mode mapping is usually applied to incorporate the largest system-bath interaction inside the system Hamiltonian and by this way allow again a perturbative treatment and possibly a Markov approximation. The reduction of the correlation timescale for the primary and secondary bath is particularly interesting in the present example. Figure 8 shows the secondary spectral density and compares the corresponding correlation functions with that of the SB model. For the SB model, the characteristic timescale is the Rabi period and for the vibronic model, the characteristic period $\tau_{EM} = 2\pi/\Omega_{EM}$ is estimated by Eq.(26).

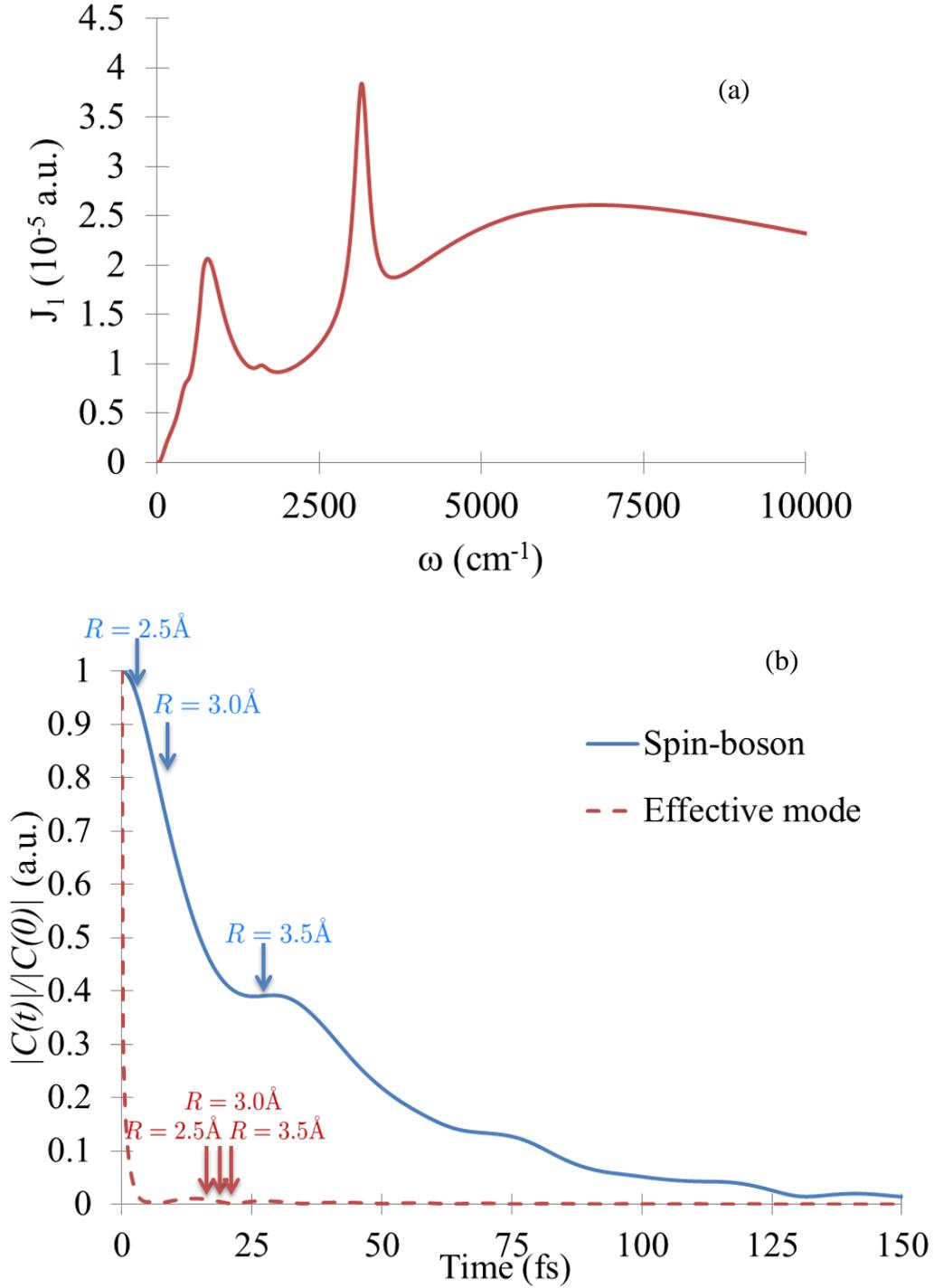

FIG. 8 Upper panel: Secondary spectral density $J_1(\omega)$ associated to the collective effective mode [Eq.(24)]. Lower panel: Normalized square modulus of the correlation function for the SB model (blue full line) and for the collective coordinate model (dashed line). The blue arrows indicate the Rabi period of the SB model and the red arrows give the characteristic period of the vibronic model estimated by Eq.(26).

When using Fig. 8, the vibronic model is expected to be more Markovian (the time scale of the system is greater than the time scale of the bath). We further use a criterion already proposed in the literature to justify a perturbative treatment. To assess its validity, following ref. [78], in Markovian, perturbative approximations, one can compare the term describing the system-bath interaction (memory kernel) with the system Hamiltonian. In the perturbative regime, the first term should be smaller than the second one, this can be expressed by a perturbative parameter $\xi_{ij}$ defined as [78]:

$$\xi_{ij} = \frac{2|S_{ij}|^2 J(\Delta E_{ij})}{e^{\beta \Delta E_{ij}} - 1} \frac{1}{\Delta E_{ij}} \ll 1. \quad (30)$$

In the spin-boson model, as it is a two-level system, this parameter is uniquely defined as $\xi_{SB}$ through the Rabi frequency ($\Delta E_{ij} = \Omega_{Rabi}$). However, for the effective mode coordinate, one has to deal with a basis set of states and not only two of them. In a first approach possible, one can take the maximum value $\xi_{max,EM}$ of this perturbation parameter $\xi_{ij}$ amongst all the transitions in this basis set :

$$\xi_{max,EM} = \max\left[\forall (i,j) \in N \times N, \xi_{ij}\right]. \quad (31)$$

This parameter is often found to be too restrictive because it depends on the basis chosen whereas not all the transitions are probed during the dynamics. We thus define another parameter that will give the validity of the perturbative method in the vicinity of the initial time which is Eq. (30) weighted by the initial populations:

$$\xi_{0,EM} = \frac{\sum_{i=1}^{N}\sum_{j=1}^{N} \rho_{S,ii}(0) \xi_{ij}(\Delta E_{ij})}{N \sum_{i=1}^{N} \rho_{S,ii}(0)} \quad (32)$$

The results are shown in Table 2.

TABLE 2. Perturbative parameters: maximum value $\xi_{max,EM}$ and weighted by the initial state one $\xi_{0,EM}$ for the effective mode model for different inter fragment distances $R$.

| $R$ (Å) | 2.50 | 3.00 | 3.50 |
|---|---|---|---|
| $\xi_{max,EM}$ | 0.318 | 0.194 | 0.145 |
| $\xi_{0,EM}$ | 0.003 | 0.001 | 0.020 |

The Markovian approximation is valid for the effective mode Hamiltonian. We can see that parameters $\xi_{max,EM}$ maximum value of this parameter amongst all the states is less than 1 but not of several orders of magnitude. However, the perturbative parameter which is weighted by the initial state $\xi_{0,EM}$ (Eq. (32)) shows that at least for short times, a perturbative treatment is possible.

We now discuss results of population obtained by the two methods. The first one is the effective mode model where the coordinate $x_1$ is exactly treated as the system coordinate and coupled to a secondary bath with a second-order perturbative Markovian master equation (Redfield equation), the second one is the reference results obtained with a spin-boson model and hierarchical equations of motion (HEOM). Results are displayed in Fig.9.

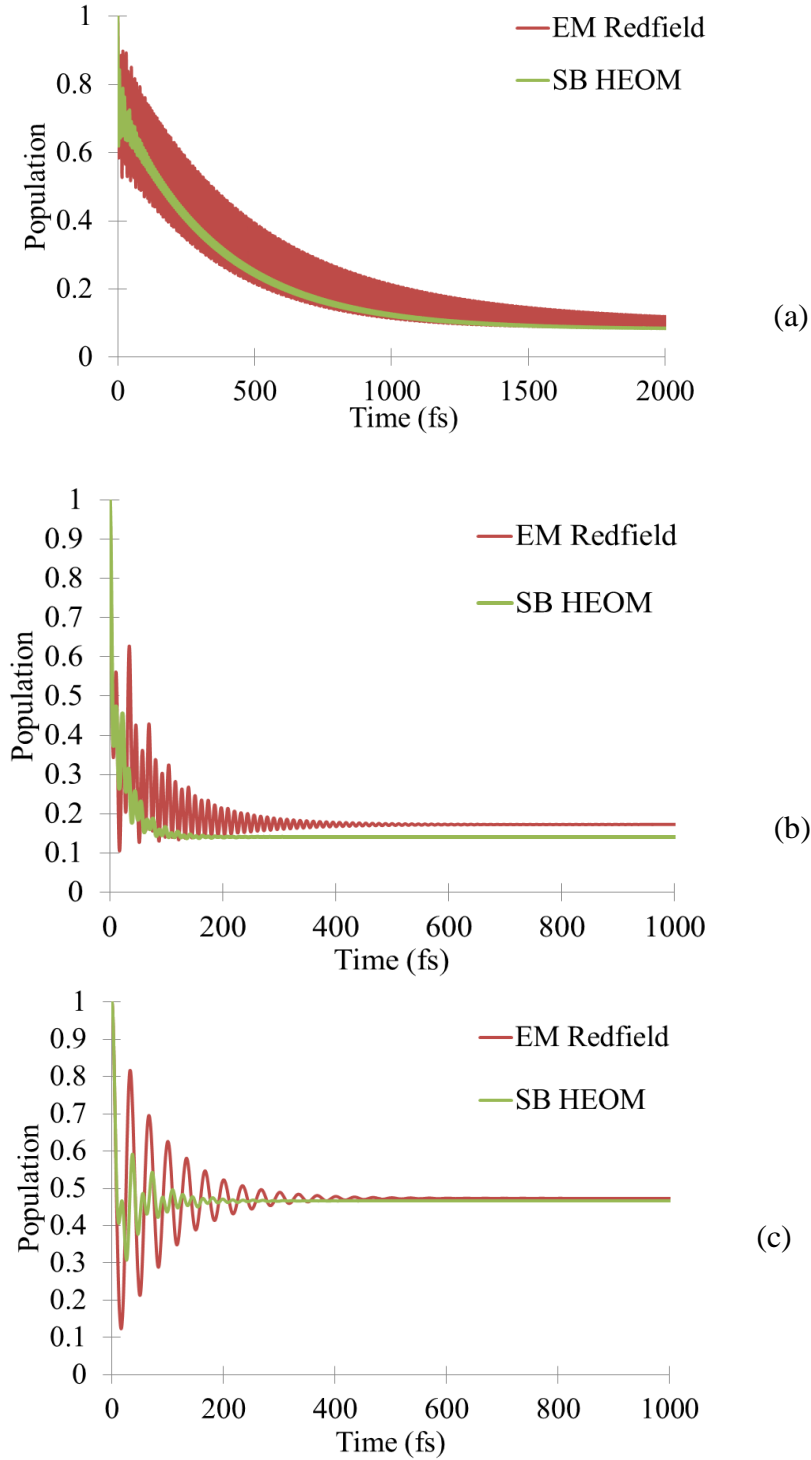

FIG. 9. Diabatic electronic populations obtained by second-order perturbation theory (Redfield) for the effective vibronic model Hamiltonian (red full line) and HEOM for the SB model (green full line). Top: $R = 2.5$ Å ; center : $R = 3.0$ Å; bottom : $R=3.5$ Å.

These results exhibit different interesting features. The Markovian effective mode Hamiltonian approach is expected to be valid as the timescale of the secondary bath correlation bath is shorter than the system timescale (the effective mode frequency here). Moreover, the perturbation parameters $\xi_{0,EM}$ at $t = 0$ are several orders of magnitude below 1 for the first two distances ($R$ = 2.5/3.0 Å) but not for the last one ($R$ = 3.5 Å). For all the distances, the $\xi_{\max,EM}$ parameters are below one but with only one order of magnitude that cannot validate the use of perturbation theory for the full dynamics. The behavior exhibited from these parameters is effectively recovered in these results. We can see that the decay of the population for the effective mode is correctly described for the first two distances even if the fine structure is not accurate. For the last distance ($R$ = 3.5 Å), the result is less satisfactory. Here, as the $\xi_{\max,EM}$ parameter is already not enough to validate perturbation theory for the whole dynamics, we can give interest to the $\xi_{0,EM}$ parameter which is worse than the one obtained for the two other distances ($R$ = 2.5/3.0 Å). Hence, discussing all these parameters (timescale of the system with timescale of the bath to validate Markov approximation, maximum value of perturbation parameter and perturbation parameter for the initial conditions) can give access to the expected quality of the subsequent dynamics.

From these considerations, we thus conclude that the methodology to single out an effective mode extracted from the spectral density to be included into the system, is a promising alternative to the full dynamics, since it might allow to revert to a perturbative and Markovian description of the secondary bath, and thus a much simpler numerical treatment than HEOM. However, its validity needs to be carefully checked for every system considered. The first favorable point is the agreement between the effective mode extracted from the spectral density and that detected in the full dynamics via the auxiliary matrices. In the present example, this main active frequency is already seen in the correlation function and corresponds to the maximum of the highest peak in the spectral density. The strategy could be less adequate in other situations with unstructured spectral density or with very many thin peals of similar intensity. Another definition of the effective coordinate must then be examined [73]. The other necessary conditions are a Markovian secondary bath and a weak residual coupling. Even though the analysis of the timescales for the secondary bath hints to a Markovian behaviour, one could in principle analyze its character using the more sophisticated measures presented in this work, applied to the vibronic system comprising ~100 states [36]. This, however, would require an exact HEOM treatment of the vibronic system, which is a very demanding task, which goes well beyond the scope of this paper.

In our case, this effective mode approach yields a qualitative correct behaviour for the populations during the transfer. In principle, one could use this approximate dynamics to analyze the non-Markovianity for the reduced electronic system. However, since the secondary bath is treated within the Markovian approximation, these measures are likely to underestimate the non-Markovianity measure.

**VII CONCLUSION**

Based on a realistic example of an electron transfer reaction accompanied by a nuclear rearrangement, we have analyzed the non-Markovian character of the dissipative process and related it to the bath dynamics. The system we consider is the oligothiophene-fullerene heterojunction, where the electron transfer between the XT and CT states depends on the distance between the $C_{60}$ and the $OT_4$ chain. In comparison with our previous work [53], we have calibrated a model using super-Ohmic Lorentzian functions for the spectral density and done exact dynamical calculations using the HEOM method. In particular, we have calculated the dissipative transfer for three distances and on the basis of these numerical results we have analyzed its dynamics.

The concept of a Markovian or non-Markovian evolution depends on the chosen system-bath separation. Once this choice is agreed upon, the ET can exhibit a non-Markovian character, as revealed by the different measures used in this work. From a physical point of view, this implies that some prominent, slow nuclear deformations affect the transfer rate. When this aspect, revealed by the measures, is further analyzed, and related to a specific nuclear motion, one can extract an effective mode to be included into an extended system. This identified mode has thus a clear physical significance. Then, understanding the close relationship between the electron transfer and the nuclear deformations for a given system might open the way to design chemical systems that exhibit particular spectral densities, which subsequently lead to enhanced or suppressed electron transfer rates.

In this context, the emphasis is laid on the non-Markovian character of the electronic process. To this end, we have chosen two recently proposed signatures of non-Markovianity, namely the canonical decoherence rates and the volume of accessible states. The non-monotonous decay of the volume during the early dynamics is connected to similar weak variations in interesting properties such as the average system energy or the entropy and therefore the free energy. According to the target, these non-Markovian effects can be a benefit or not. This opens the perspective of control of these effects by changing the surrounding either by interaction with an electric field of by chemical engineering. In this context, the experimental determination of the volume or of signatures of non-Markovianity in chemical systems seems to be crucial even if probably difficult. In principle, the volume could be obtained by quantum tomography by preparing and analyzing the three initial states to compute the dynamical map [36]. A tomography protocol has been discussed in the context of energy transfer in condensed phase by non-linear spectroscopy [79, 80]. Detection of experimental signals such as the electric field generated by the charge separation at heterojunction could also give information about non-monotonous evolution [81].

We have compared the two measures discussed in this work with an *a priori* analysis of the system and bath timescales involved. As main findings, we have shown that the non-Markovianity measure based on the volume of accessible states is in good agreement with the estimate based on system/bath timescales. As for the canonical decoherence rates, which gives a much more detailed view, we have found in some cases negative rates, even though the volume measure decreases at any time. In the present three examples, the non Markovian measure based on the volume evolves as predicted by the *a priori* estimation based on the system and bath timescales.

Additionally, an interesting result is the connection between the oscillatory pattern of the canonical rates and the fundamental frequency of the damped bath collective coordinate in both electronic states. This information on the first moment of the collective mode dynamics is provided by the first level auxiliary matrices within the HEOM formalism. Furthermore, in this example a very good agreement has been obtained with the effective mode extracted from the spectral density derived by a strategy already proposed in ref. [75]. Within this approach, the system is augmented by a collective coordinate, which is coupled to a secondary bath. A part of the non-Markovianity can thus be captured and integrated into the augmented systems dynamics. This result emphasizes the close relationship between the non-Markovianity and the choice of the system-bath partition as we were discussing it earlier in this conclusion. This procedure could be very interesting for further investigation for instance for control simulations.[82] The nice agreement between the HEOM mode and the effective mode extracted from the spectral density is due to its particular shape clearly dominated by a single peak indicating the main damping mode. For spectral density with large background or for highly structured density with peaks of similar heights the efficiency of this procedure implying a single mode is not guaranteed and an investigation about the best effective mode remains necessary. The second caveat concerns the Markovian treatment which also requires a careful justification.

As a perspective, we plan to study the same non-Markovian measures with an effective mode but by performing the dynamics with other levels of theory. However, to describe the secondary bath at an exact level, one needs to extend the HEOM formalism to include a coordinate. Work along this line is currently being pursued.

These considerations of the canonical rates and corresponding decay channels are of great importance in the context of control by electric fields of complex molecular systems exhibiting non-Markovian behavior so that control and dissipation are strongly correlated. The frequency of the bath mode connected to the flow back from the environment gives an indication of the period where Stark shift induced by the electric field could modify the system-bath coupling in order to enhance or remove the non-Markovianity and its effect on important properties such as average energy or entropy.

**ACKNOWLEDGMENTS**

We would like to thank Prof. I. Burghardt for providing the data necessary to calibrate the spin-boson model of the heterojunction.

We also acknowledge Prof. D. Sugny, Dr O. Atabek and Dr R. Puthumpally-Joseph for very fruitful discussions. E.M. thanks Prof. Q. Shi for visiting his team in Beijing and providing a grant.

We acknowledge support from the French and German science foundations through project DFG-ANR, under grant N0 ANR-15-CE30-0023-01. This work has been performed within the French GDR 3575 THEMS

**APPENDIX**

**A. Spectral density and correlation function**

The parameters of the four-pole Lorentzian functions fitting the spectral density (see Eq.(4)) are as follows :

| $l$ | 1 | 2 | 3 | 4 |
|---|---|---|---|---|
| $p_l$ | $1.900\ 10^{-17}$ | $9.077\ 10^{-15}$ | $4.473\ 10^{-13}$ | $1.303\ 10^{-15}$ |
| $\Omega_{l,1}$ | $1.934\ 10^{-03}$ | $6.993\ 10^{-03}$ | $7.397\ 10^{-03}$ | $1.446\ 10^{-02}$ |
| $\Gamma_{l,1}$ | $5.551\ 10^{-04}$ | $5.463\ 10^{-04}$ | $5.652\ 10^{-04}$ | $6.039\ 10^{-04}$ |
| $\Omega_{l,2}$ | $1.020\ 10^{-04}$ | $2.729\ 10^{-03}$ | $1.204\ 10^{-02}$ | $3.077\ 10^{-03}$ |
| $\Gamma_{l,2}$ | $6.120\ 10^{-04}$ | $8.378\ 10^{-04}$ | $1.80\ 10^{-02}$ | $2.03\ 10^{-04}$ |

The integral (6) giving the bath correlation function is performed analytically with the parametrization (4) of the spectral density. The residue theorem is used to compute the integral with the contour over the upper half-plane enclosing $4n_l$ poles in $(\Omega_{l,1}, \Gamma_{l,1})$, $(-\Omega_{l,1}, \Gamma_{l,1})$, $(\Omega_{l,2}, \Gamma_{l,2})$, $(-\Omega_{l,2}, \Gamma_{l,2})$ and an infinity of poles on the imaginary axis $\left\{\forall j \in \mathbb{N}^* / \left(0, v_j = \frac{2\pi}{\beta} j\right)\right\}$ called the Matsubara frequencies. As these poles are of order 1, the parameters $\alpha_k$ and $\gamma_k$ can be expressed as:

$$\gamma_{l,1} = \Omega_{l,1} + i\Gamma_{l,1} \tag{A1}$$

$$\gamma_{l,2} = -\Omega_{l,1} + i\Gamma_{l,1} \tag{A2}$$

$$\gamma_{l,3} = \Omega_{l,2} + i\Gamma_{l,2} \tag{A3}$$

$$\gamma_{l,4} = -\Omega_{l,2} + i\Gamma_{l,2} \tag{A4}$$

$$\gamma_{j,matsu} = iv_j \tag{A5}$$

$$\alpha_{l,1} = \frac{p_l \gamma_{l,1}^2}{8\Omega_{l,1}\Gamma_{l,1}\left[\left((\gamma_{l,1} - \Omega_{l,2})^2 + \Gamma_{l,2}^2\right)\left((\gamma_{l,1} + \Omega_{l,2})^2 + \Gamma_{l,2}^2\right)\right]} \left(\coth\left[\frac{\beta}{2}\gamma_{l,1}\right] - 1\right) \tag{A6}$$

$$\alpha_{l,2} = \frac{p_l \gamma_{l,2}^2}{8\Omega_{l,1}\Gamma_{l,1}\left[\left((\gamma_{l,2} - \Omega_{l,2})^2 + \Gamma_{l,2}^2\right)\left((\gamma_{l,2} + \Omega_{l,2})^2 + \Gamma_{l,2}^2\right)\right]} \left(\coth\left[-\frac{\beta}{2}\gamma_{l,2}\right] + 1\right) \tag{A7}$$

$$\alpha_{l,3} = \frac{p_l \gamma_{l,3}^2}{8\Omega_{l,2}\Gamma_{l,2}\left[\left((\gamma_{l,3} - \Omega_{l,1})^2 + \Gamma_{l,1}^2\right)\left((\gamma_{l,3} + \Omega_{l,1})^2 + \Gamma_{l,1}^2\right)\right]} \left(\coth\left[\frac{\beta}{2}\gamma_{l,3}\right] - 1\right) \tag{A8}$$

$$\alpha_{l,4} = \frac{p_l \gamma_{l,4}^2}{8\Omega_{l,2}\Gamma_{l,2}\left[\left((\gamma_{l,4} - \Omega_{l,1})^2 + \Gamma_{l,1}^2\right)\left((\gamma_{l,4} + \Omega_{l,1})^2 + \Gamma_{l,1}^2\right)\right]} \left(\coth\left[-\frac{\beta}{2}\gamma_{l,4}\right] + 1\right) \tag{A9}$$

$$\alpha_{j,matsu} = \frac{2i}{\beta} J(iv_j) \tag{A10}$$

**B. Decoherence matrix**

A numerical strategy to compute the decoherence matrix is given by the expression [31]

$$D_{ij}(t) = \sum_{m=0}^{d^2-1} Tr\left[G_m G_i \Lambda_t [G_m] G_j\right] \tag{A11}$$

The signification of $\Lambda_t[G_m]$ is clear by recasting of the master equation. From Eq.(11) one has

$$\rho(t) = \sum_{k,l=0}^{d^2-1} F_{lk}(t) Tr(G_k \rho(0)) G_l \tag{A12}$$

and the corresponding master equation reads

$$\dot{\rho}(t) = \sum_{k,l=0}^{d^2-1} \dot{F}_{lk}(t) Tr(G_k \rho(0)) G_l. \quad (A13)$$

If the dynamical map is invertible, one has also

$$\dot{\rho}(t) = \sum_{k,l=0}^{d^2-1} \dot{F}_{lk}(t) Tr(G_k \phi_t^{-1}[\rho(t)]) G_l. \quad (A14)$$

By expanding again $\rho(t) = \sum_{j=0}^{d^2-1} Tr(G_j \rho(t)) G_j$ and by introducing the matrix representation of the inverse map $F_{kj}^{-1} = Tr(G_k \phi_t^{-1}[G_j])$, one gets

$$\begin{aligned}\dot{\rho}(t) &= \sum_{l,j=0}^{d^2-1} \left(\dot{F}(t) F^{-1}(t)\right)_{lj} Tr(G_j \rho(t)) G_l \\ &= \sum_{l,j=0}^{d^2-1} M_{lj}(t) Tr(G_j \rho(t)) G_l\end{aligned} \quad (A15)$$

The $M(t) = \dot{F}(t) F^{-1}(t)$ matrix is the logarithmic derivative of the dynamical map and can be written $M_{lj}(t) = Tr(G_l \Lambda_t [G_j])$ with

$$\Lambda_t[G_j] = \sum_{k=0}^{d^2-1} \dot{\phi}_t[G_k] F_{kj}^{-1} \quad (A16)$$